\documentclass[twocolumn,showpacs,prl,floatfix,eps]{revtex4}
\ifx\pdfoutput\undefined
\usepackage{graphicx}
\else
\usepackage{epstopdf}
\usepackage{epsfig}
\usepackage{color}

\usepackage{amssymb,amsmath,stmaryrd,tabularx}
\usepackage{wrapfig}
\usepackage{bm}
\fi

\usepackage[center]{subfigure}

\newcommand{\be}{\begin{equation}}
\newcommand{\ee}{\end{equation}}
\newcommand{\bea}{\begin{eqnarray}}
\newcommand{\eea}{\end{eqnarray}}

\begin{document}
\title{Intermittent dislocation density fluctuations in crystal plasticity from a phase-field crystal model}
\author{Jens M. Tarp$^1$, Luiza Angheluta$^{2}$,  Joachim Mathiesen$^1$, and Nigel Goldenfeld$^{3}$}
\affiliation{
$^1$ Niels Bohr Institute, University of Copenhagen, Blegdamsvej 17, DK-2100 Copenhagen, Denmark. \\
$^2$ Physics of Geological Processes, Department of Physics, University of Oslo, P.O. 1048 Blindern, 0316 Oslo
Norway\\
$^3$Department of Physics, University of Illinois at
Urbana-Champaign, Loomis Laboratory of Physics, 1110 West Green
Street, Urbana, Illinois, 61801-3080}


\begin{abstract}
Plastic deformation mediated by collective dislocation dynamics is
investigated in the two-dimensional phase-field crystal model of
sheared single crystals. We find that intermittent fluctuations in the
dislocation population number accompany bursts in the plastic
strain-rate fluctuations. Dislocation number fluctuations exhibit a
power-law spectral density $1/f^2$ at high frequencies $f$. The
probability distribution of number fluctuations becomes bimodal at low
driving rates corresponding to a scenario where low density of defects
alternate at irregular times with high population of defects. We
propose a simple stochastic model of dislocation reaction kinetics that
is able to capture these statistical properties of the dislocation
density fluctuations as a function of shear rate.
\end{abstract}
\maketitle
\date{}

Small-scale plasticity is characterized by intermittent strain-rate
fluctuations and strain avalanches that follow robust power-law
statistics, with detailed reports for both crystalline materials
~\cite{richeton2005dislocation,richeton2006critical,greer2009situ} and
amorphous  materials~\cite{argon2013strain}. Discrete dislocation
dynamics simulations have been successful in reproducing the power-law
statistics of strain avalanches in crystal plasticity assuming that the
microscopic origin of intermittency is attributed to dissipative
processes associated with crystal defects, such as dislocations and
disclinations~\cite{miguel2001intermittent,ispanovity2010submicron,
tsekenis2013determination}. The scaling behavior observed near plastic yielding was initially explained by analogy to the depinning phase transition~\cite{friedman2012statistics,tsekenis2013determination,leblanc2012distribution}. However, recent numerical simulations and experiments show that a scale free behavior of plastic bursts occurs also at applied stresses away from the yielding point, which makes the relation to nonequilibrium depinning transition a controversional topic~\cite{ispanovity2014avalanches,ispanovity2013average}.

As the global plastic strain rate, $\dot\gamma$, is directly
proportional to the mobile dislocation density $\rho_d$ and mean velocity
$\langle v\rangle$  by Orowan's relation: $\dot \gamma\approx b \rho_d
\langle v\rangle $, the intermittency of $\dot\gamma$ is influenced
both by the collective dislocation velocity (typically following
stick-slip dynamics) and dislocation number fluctuations.  Previous
work on plastic avalanches has only taken into account the stick-slip
motion of crystal defects, due to the challenging nature of modeling dislocation density fluctuations naturally, without
introducing artifacts due to ad hoc rules for dislocation reactions.
The question that concerns us here is: what is the effect of
dislocation number fluctuations, particularly near the yielding
transition?

The purpose of this Letter is to investigate numerically the density
fluctuations of mobile dislocations as a crystal is sheared near and away
from yielding point. Conventional techniques that solve the
equations of motion for discrete dislocations are not appropriate,
because they would require ad hoc rules for dislocation annihilation
and creation.  Instead, we use the phase-field crystal
approach~\cite{Elder04} which has been shown to be an efficient
technique to model plastic deformations in single and poly-crystals,
one which can capture implicitly dislocation dynamics and interactions~\cite{emmerich2012phase,berry2014phase}. Working in two
dimensions, we find that the total number of defects $N_d$ is a
highly fluctuating quantity whose power spectrum at large frequency $f$ is
characterized by a $1/f^2$ scaling, similar to that of the
global plastic strain rate. Also, $N_d$ follows a non-trivial
probability distribution which becomes bimodal at small driving rates,
reflecting that both dislocation extinction events and a dense
population of interacting dislocations occur with a non-zero
probability. Finally, we develop a stochastic coarse-grained model for
the effective continuum mechanics of our phase-field crystal model,
based upon previous work that has successfully described the regime of
cyclic loading and plastic regime where dislocations self-organize into
cell-like
patterns~\cite{hahner1996foundations,hahner1996stochastic,ananthakrishna2007current}.
In this way, we are able to calculate the power spectrum of dislocation
number fluctuations, obtaining results in agreement with our
simulations. While we have not performed extensive tests in three dimensional systems, our stochastic model seems to remain valid despite the more complex defect structures admissible in three dimensions.

{\it Sheared phase-field crystal:-} We use the phase-field crystal
(PFC) model to study intermittent plastic deformation in single
crystals. The PFC model describes the evolution of an order parameter
field $\psi(\bm r,t)$ that is related to the particle number density
averaged over atomic vibration timescales, i.e. $\psi\sim \langle
\sum_i\delta(\bm r-\bm r_i)\rangle_t$. An effective Swift-Hohenberg
free-energy functional is formulated in the lowest order gradient
expansion of $\psi$-field and given as~\cite{Elder04}
$\mathcal{F}\{\psi\} = \int d\bm r
\left[\frac{1}{2}\psi\left(q_0^2+\nabla^2\right)^2\psi+\frac{r}{2}\psi^2+\frac{1}{4}\psi^4\right]$,
where $r\sim (T-T_c)/T_c<0$ is the quenching depth parameter related to
the deviation from the critical melting temperature, and $q_0 = 2\pi/a$
with $a$ being the equilibrium lattice spacing. The equilibrium phase
diagram obtained from the free energy $\mathcal{F}\{\psi\}$ contains
a region in the $(r,\psi_0)$-space where the system relaxes to a spatially periodic $\psi$-field around a mean crystal density $\psi_0$, that corresponds to a crystalline phase with triangular symmetry in two-dimensions~\cite{Elder04}.

The dynamics of the $\psi$-field that includes both the diffusive
timescale of phase transformation and elastic strain relaxation is
given by a damped wave equation~\cite{stefanovic2006phase}
\begin{equation}\label{eq:PFC}
\frac{\partial^2\psi}{\partial t^2}+\beta \frac{\partial \psi}{\partial t}= -\nabla\cdot\bm J,
\end{equation}
where the density current has a diffusive part determined by the free
energy $\mathcal{F}$ and an advective part that simulates the strain rate boundary conditions, namely
%
$\bm J = -\alpha^2\nabla\frac{\delta\mathcal{F}}{\delta \psi}+\beta\bm v\psi$,
%
where $\alpha^2$ is the diffusivity coefficient and $\beta$ is the overdamped coefficient. Physically, these two parameters are related to the effective sound speed and vacancy diffusion coefficient that set a finite elastic interaction length $L^*$ and time $t^*$~\cite{Stefanovic09}. Their values are constrained by the thermodynamic stability of the crystalline phase and the damping rate of the elastic excitations. From a linear stability analysis around the one-mode approximation solution of
$\psi$, the dispersion relation corresponding to Eq.~(\ref{eq:PFC}) with $\bm v=0$ describes a pair of density waves that propagate undamped
for a time $t^*\approx 2\beta^{-1}$ over a lengthscale $L^*\approx
v_{eff}t^*$, with an effective wave speed $v_{eff}\approx
2\alpha\sqrt{3\psi_0^2+r+q_0^4+9A_0^2/8}$, where $A_0 =
4/5(\psi_0+1/3\sqrt{-15r-36\psi_0^2})$ is the amplitude of $\psi$ in
the one-mode approximation~\cite{Stefanovic09}. For timescales $t>t^*$,
the density disturbance propagates diffusively with a vacancy diffusion
coefficient $D_0= \alpha^2(3\rho^2+r+q_0^4+9A_0^2/8)/\beta$. The model
parameters are chosen such that the timescale $t^*$ of strain wave
relaxation is much smaller than the diffusion one. We impose a constant strain rate boundary condition similar to that used in Ref.~\cite{Chan10}. The velocity field $\bm v = (v_x(y),0)$
acts effectively only on a small strip on the top and bottom boundaries,
\begin{equation}
v_x(y) = v_0 e^{-\frac{L-y}{\lambda}} H\left(y-\frac{L}{2}\right)-v_0 e^{-\frac{y}{\lambda}} H\left(\frac{L}{2}-y\right),
\end{equation}
where $H(x)$ is the Heaviside step function and the penetration length
$\lambda$ of the imposed shear velocity is chosen to be much smaller than the system
size $L$.

We solve Eq.~(\ref{eq:PFC}) numerically on a two-dimensional
rectangular $L\times L$ domain of size $512 dx\times 512 dx$, using both a
finite difference method with an isotropic discretization of the gradients
and the  Laplace operator~\cite{PRE12} as well as a spectral method
utilizing discrete cosine transforms similar to that in Ref.~\cite{tegze2009advanced}. In the spectral method implementation, we added a conservative, Gaussian noise term to Eq.~(\ref{eq:PFC}) to enable dislocation nucleation in the regime of low dislocation density. Regardless of the noise term or numerical implementation, we observe robust statistical properties. The discretization parameters are
set to $dt = 0.015$, $dx=1$, and $a=2\pi$. The boundary conditions are
periodic in the $x$- direction and with zero-flux in the $y$-direction
at $0$ and $L$. As initial condition, we use the relaxed equilibrium
solution of a single crystal at a fixed undercooling depth $r= -0.5$
and $\psi_0 = 0.3$. The damping coefficient is set
to $\beta = 0.5$ and the diffusivity parameter $\alpha=15$. These
parameter values give a typical wave speed of $v_{eff}\approx 36$, an
elastic interaction length $L^*\approx 146$, and a characteristic
damping time $t^*\approx 4$. We set $\lambda = 0.05 L <L^*$, thus the
strain waves propagate deeper into the bulk of the crystal before they
are dissipated. However, in a constant strain-rate experiment, the
imposed shear rate should be sufficiently slow such that the elastic
waves decay almost~\lq instantaneously\rq. Under these conditions, we
vary the imposed shear velocity between $v_0 \approx 4.3-8.4$, that
corresponds to an applied strain rate varying between $\dot\epsilon_0 =
\lambda v_0/(2L^2) \approx 2.10-4.10\times 10^{-4}$ in arbitrary time
units.

\begin{figure}[t]
\begin{centering}
 (a) \hspace{15 mm}      (b)\\
 \includegraphics[width=0.45\columnwidth]{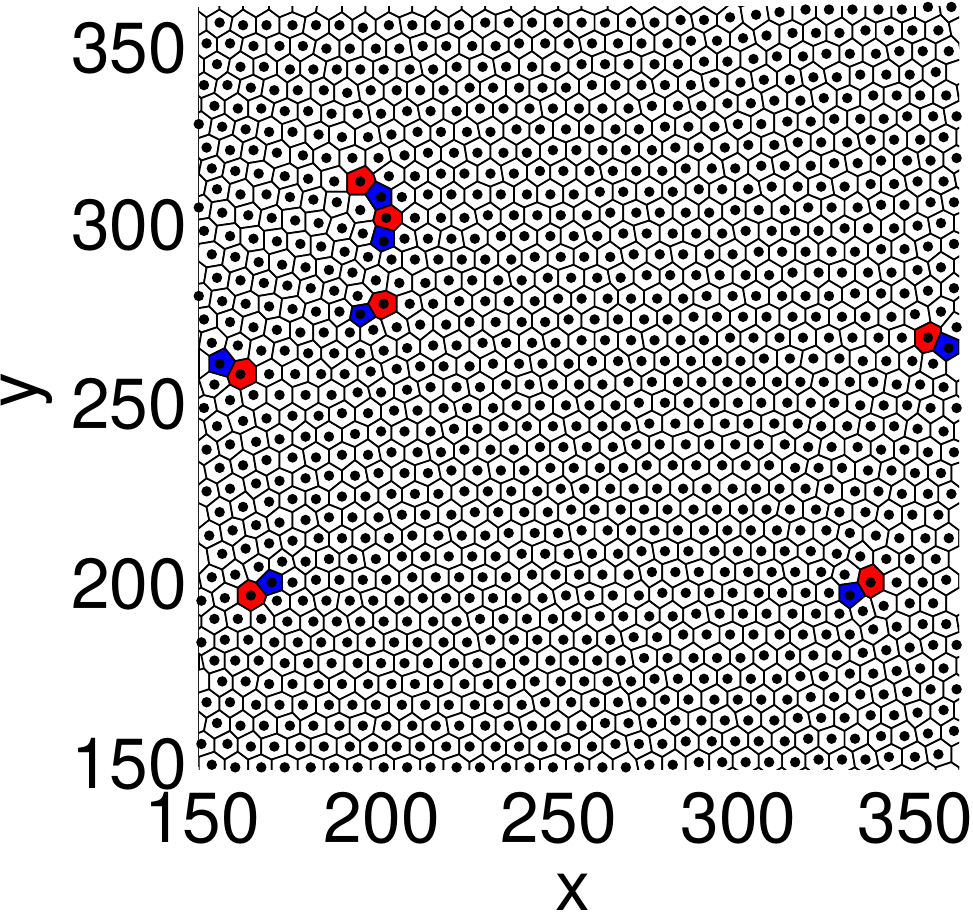}
  \includegraphics[width=0.45\columnwidth]{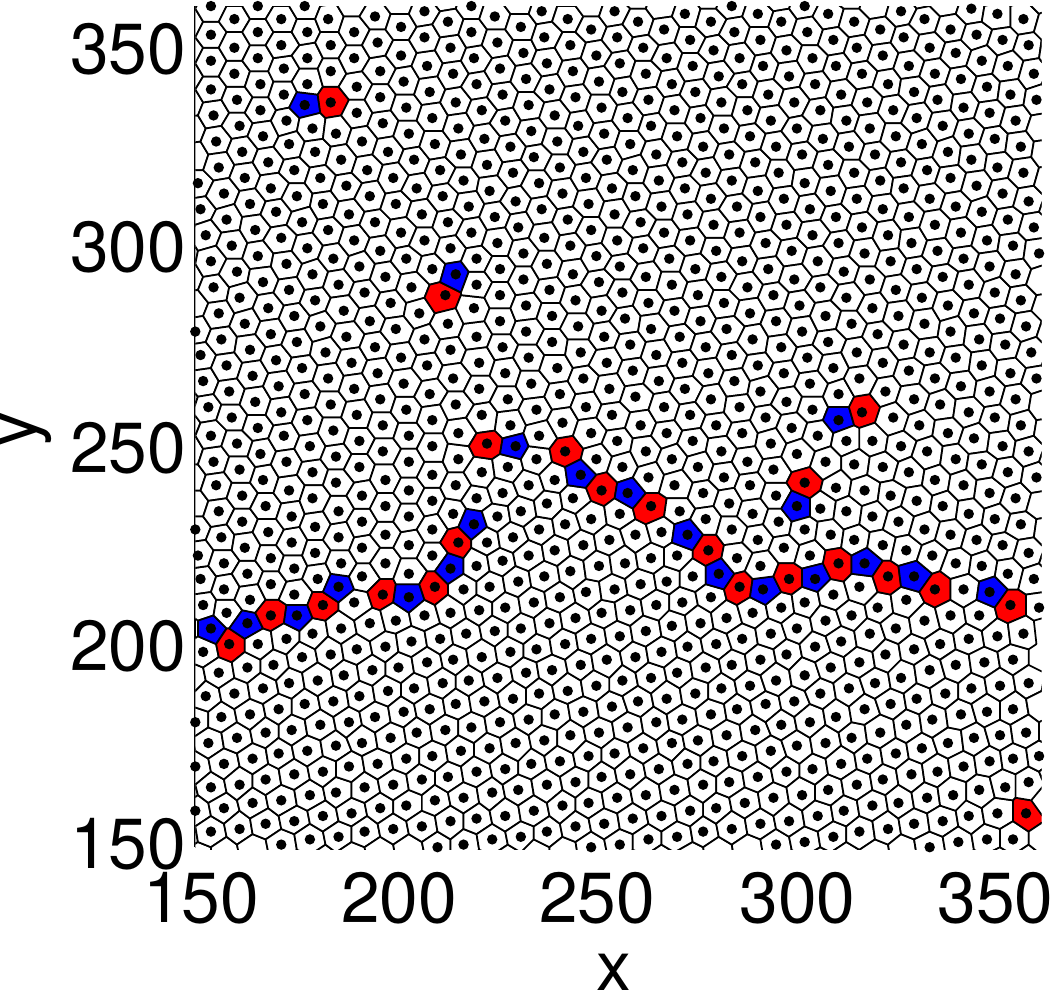}\\
  (c) \hspace{15 mm}      (d)\\
  \includegraphics[width=0.42\columnwidth]{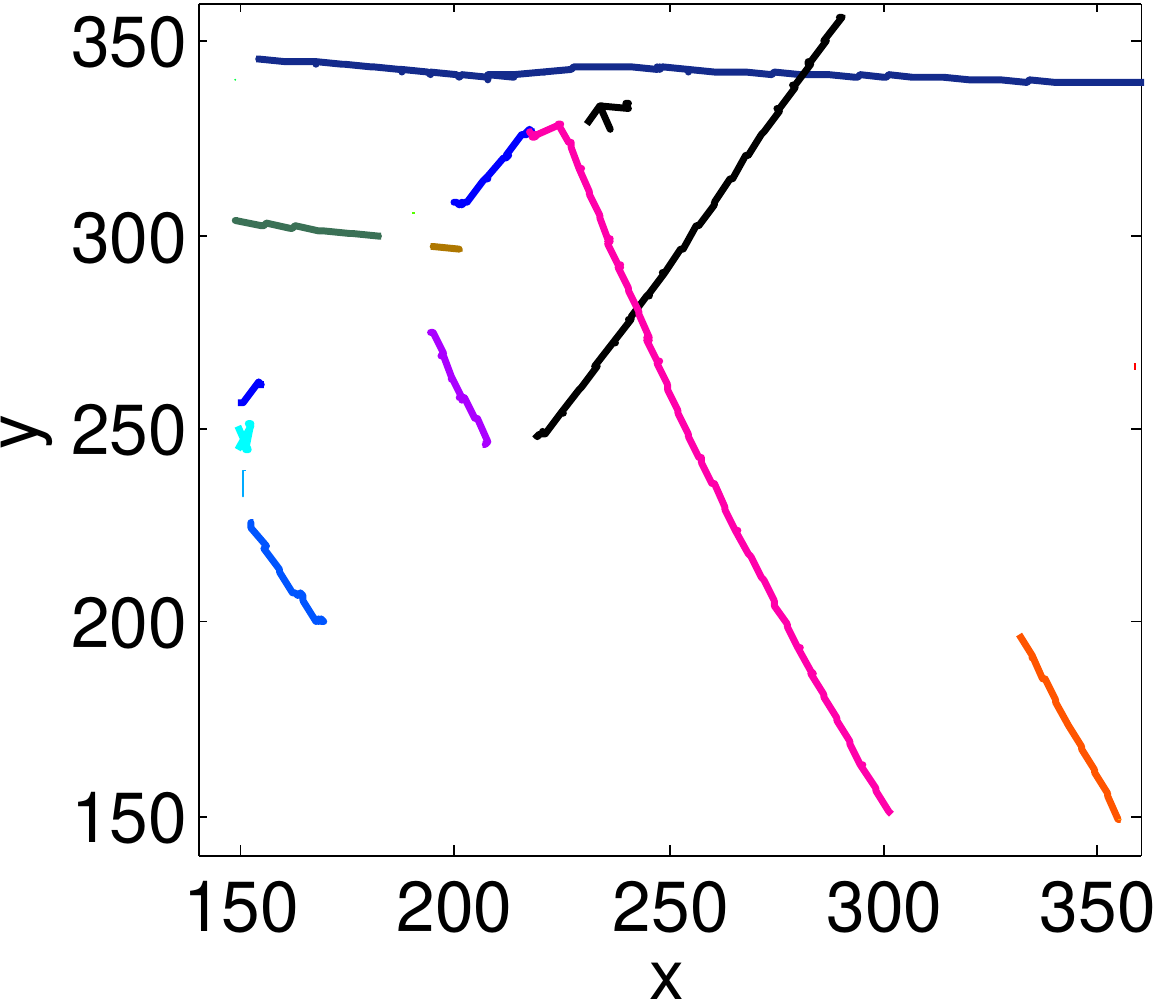}
  \includegraphics[width=0.45\columnwidth]{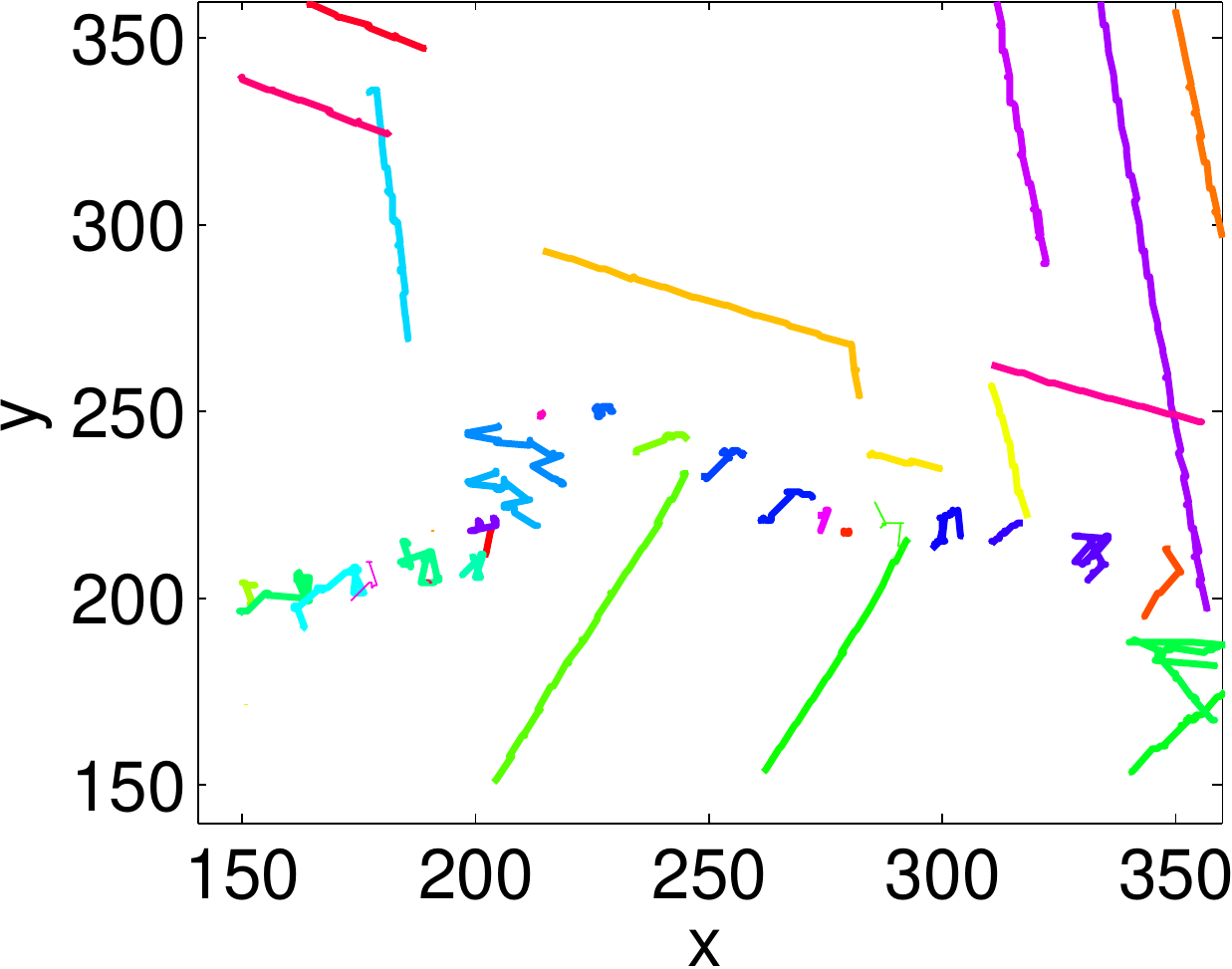}
  \caption{\label{fig:2} (Color online) Different dislocation configurations in the Voronoi diagram (panels (a) and (b)) corresponding to a crystal that is sheared  at constant rate $\dot\epsilon_0 = 3\times 10^{-4} s^{-1}$. Dislocations are represented as pairs of 5-fold and 7-fold defect atoms shown in red and blue. The trajectories of these dislocations are depicted (random coloring) in the bottom panels, (c) for a dilute configuration of dislocations and (d) for a more dense distribution of dislocations organized into a wall sweeping across the system.}
\end{centering}
\end{figure}

{\it Dislocation density fluctuations:-} Plastic deformations induce
vacancies and point topological defects that are represented by phase
and amplitude modulations in the $\psi$-field. These defects are harder
to locate and track accurately from the crystal density field. Instead,
we use a more atomistic approach of tracking the atoms neighboring
different point defects. For a perfect fcc lattice in two dimensions, each PFC atom (located as a minimum in the
$\psi$-field) is surrounded by $6$ nearest neighbors, but
near a point defect the coordination number is different. Defect atoms
with coordination number $5$ or $7$ are associated with
disclinations, and typically pair up to form a dislocation as illustrated in Figure~(\ref{fig:2}).
\begin{figure}[t]
\begin{centering}
 \includegraphics[width=\columnwidth]{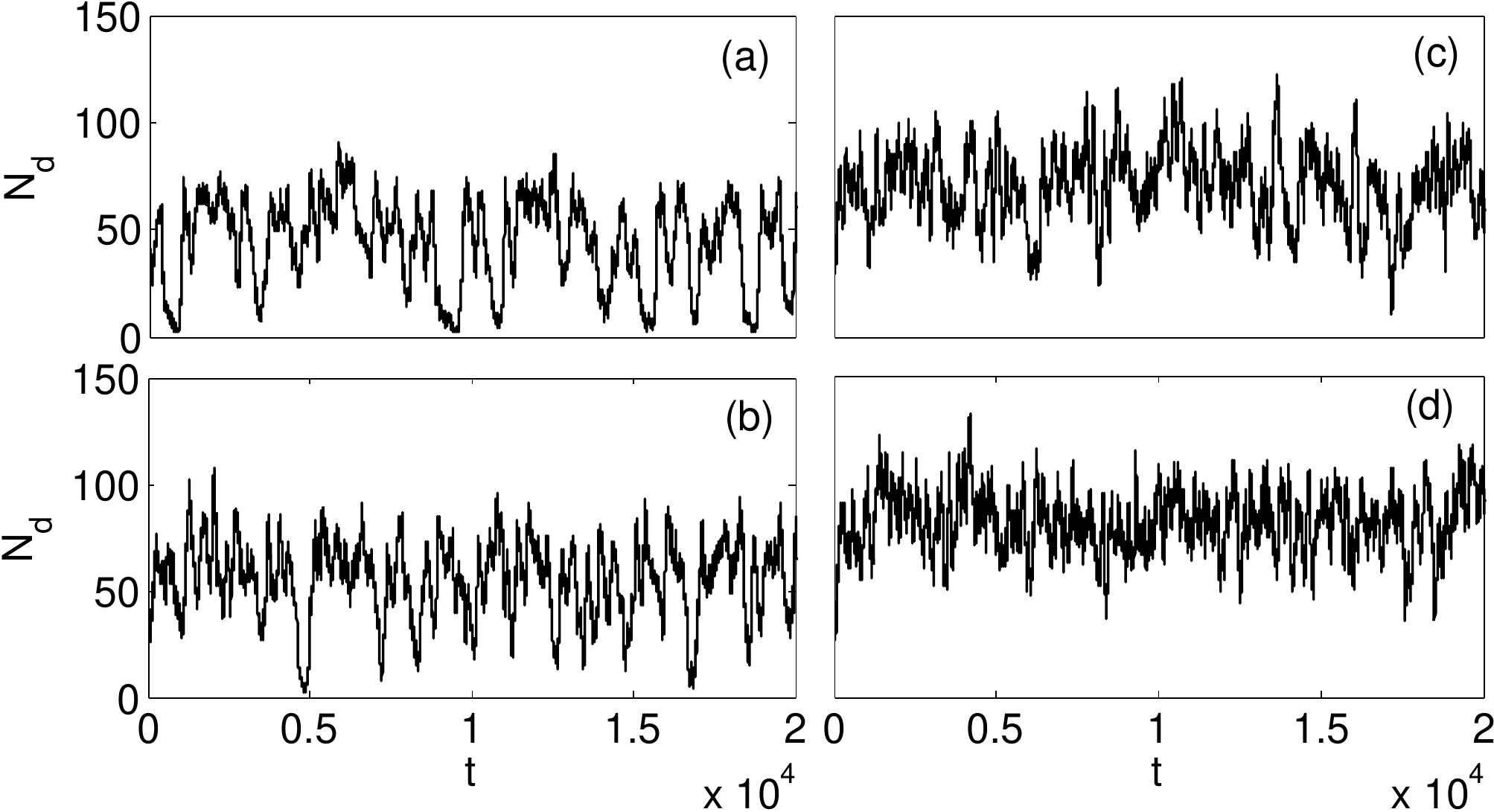}
\end{centering}\caption{\label{fig:3} (Color online) Temporal
fluctuations in the dislocation number for different strain rates. From
(a) to (d) the strain rate increases as $2.78\times 10^{-4}$,
$3.02\times 10^{-4}$, $3.41\times 10^{4}$, and $3.90\times 10^{-4}$.}
\end{figure}

We focus on the interaction and dynamics of dislocations
which are defined as pairs of $5$ and $7$-fold defect atoms. The number
of dislocations varies with the system size and the applied strain
rate, and, for our parameter values, it is up to $\mathcal{O}(10^2)$.
For a given strain rate, the dislocation density is both
highly heterogeneous in space and intermittent in time. The spatial density
of dislocations varies from dilute configuration of single, fast-moving
dislocations to denser configurations of dislocation pile-ups that
slowly sweep across the sheared crystal (see Figure~(\ref{fig:2}) (a)-(b)). The
trajectories of isolated dislocations follow the slip planes at fast
speeds, in contrast to
the slow motion of correlated dislocations (see Figure~(\ref{fig:2}) (c)-(d)). The alternations between
configurations of fast and slow moving dislocations result in
sporadic fluctuations of the plastic strain rate that is related to the
global dislocation velocity. Ref.~\cite{Chan10} studied the avalanche
statistics of the global strain rate signal showing that the power-law
exponent of the energy dissipated during avalanches is in agreement
with that predicted near a depinning transition in the mean field
approximation. Here we study instead the nontrivial statistical
properties of the dislocation density fluctuations. A detail analysis
of the relation between the statistics of number fluctuations and a
depinning phase transition is the subject of a separate study.

\begin{figure}[t]
\begin{centering}
 \includegraphics[width=0.8\columnwidth]{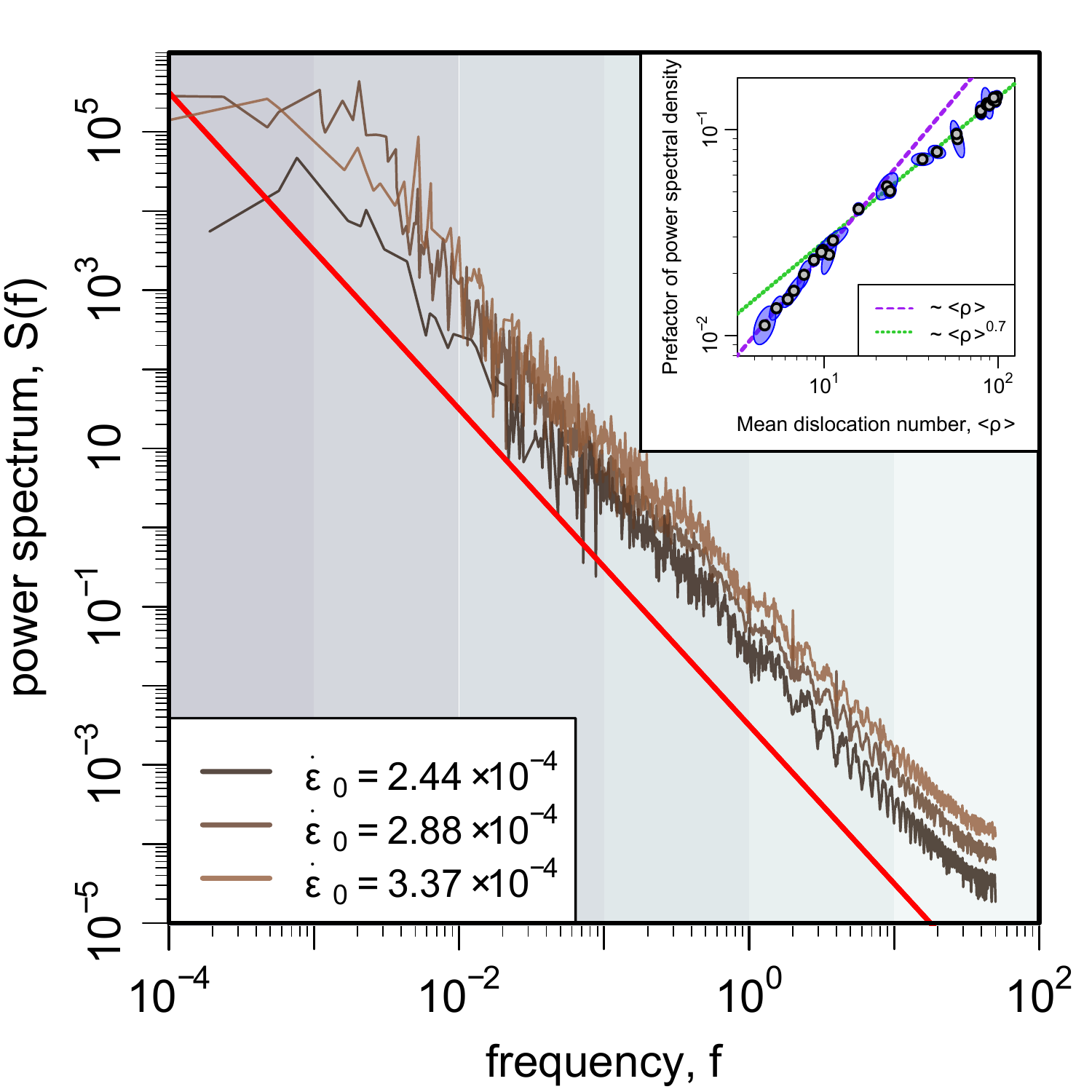}
\end{centering}\caption{\label{fig:4} (Color online) Power spectrum of
the dislocation number fluctuations $N_d$ vs. frequency $f$ for
different strain rates. The red line shows the prediction of the stochastic model
Eq. (\ref{eq:Lagevin}), varying as $C/f^2$ for large $f$. The inset shows the dependence of $C$ on the mean dislocation number. The error bars in the inset are given along the principal axes of the variation over multiple simulations of the mean of $N_d$ and $C$. }
\end{figure}

We observe that sudden dislocation reactions (annihilations or
creations) occur at irregular times between isolated dislocations or
between an isolated dislocation and a domain wall. The latter event may
lead to the breaking-up of the domain wall and the release of
fast-moving dislocations along different gliding planes. The total
dislocation number $N_d$ is a highly fluctuating quantity depending on
the imposed shear rate, such that, at low strain rates, it is
characterized by temporal variations around a mean set by the external
driving interspersed with short episodes of almost dislocation
extinction, as seen in Figure~(\ref{fig:3}). For $\dot\epsilon_0<2.00 \times 10^{-4}$, keeping the same values for the
other parameters, we observe that the vanishing dislocation density
becomes an absorbing state after an initial transient time of
fluctuations. In this dynamical regime, the crystal has been rotated
such that the stored energy is mainly dissipated by visco-elastic
deformation without the nucleation of defects.

In Figure~(\ref{fig:4}), we show the power spectrum $S(f) =
\langle|\hat N_d|^2\rangle$, where $\hat N_d$ is the Fourier transform
of $N_d$, computed from the time signals illustrated in
Figure~(\ref{fig:3}). The power spectrum has a power-law decay at high
frequencies $f$ given by $C/f^2$. To test that this scaling behavior arises of the density fluctuations from correlated events, we have measured the dependence of the scaling coefficient $C$ on $\langle  N_d \rangle$ shown in the inset of Figure~(\ref{fig:4}). For low shear rates corresponding a dilute dislocation density, the coefficient exhibits a linear scaling, that is
consistent with uncorrelated, random dislocation reactions. At higher shear rates, the dependence changes to a power-law implying that the signal arises from correlated dislocation dynamics.

\begin{figure}[t]
\begin{centering}
 \includegraphics[width=0.8\columnwidth]{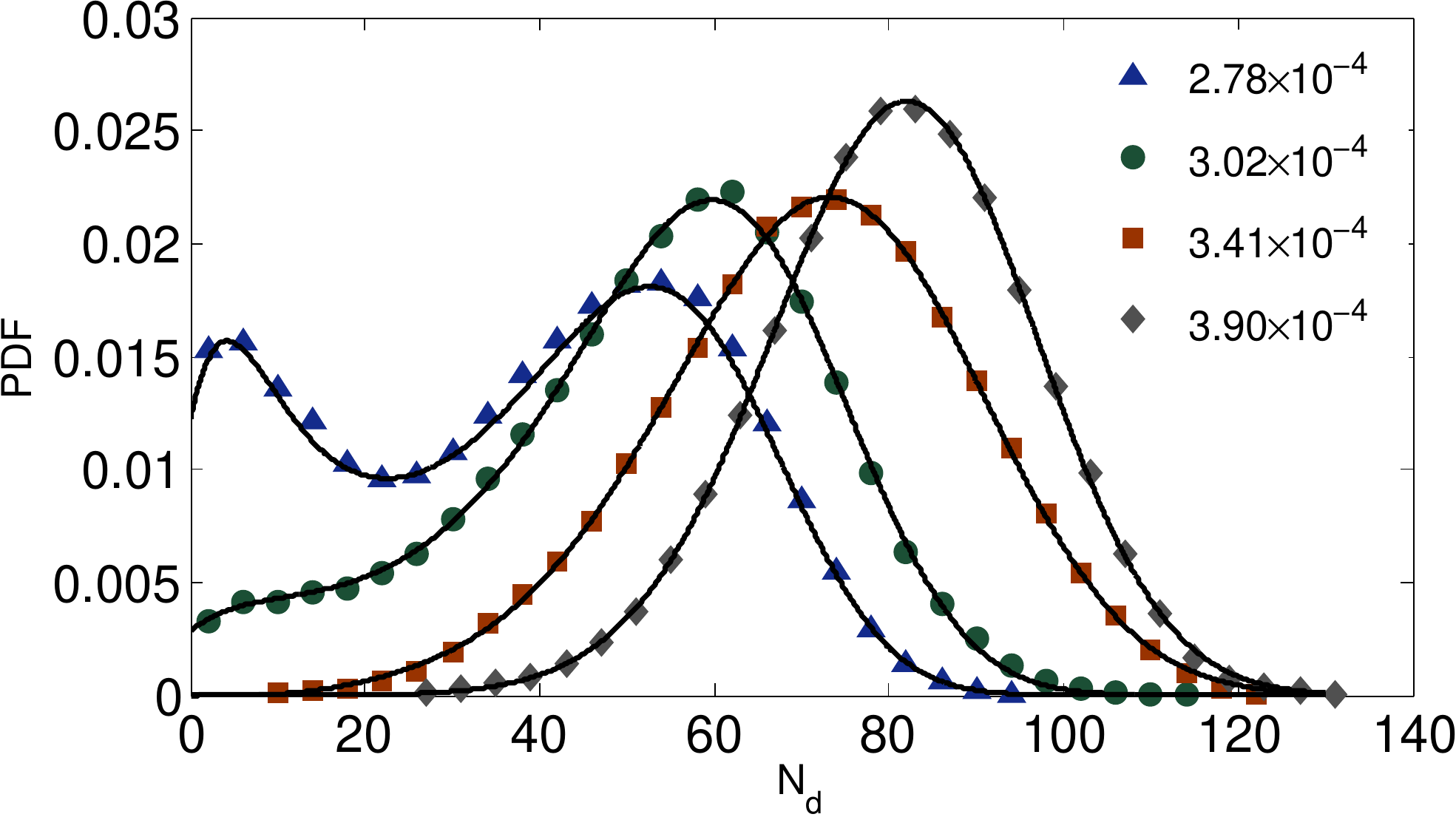}
\end{centering}\caption{\label{fig:5} (Color online) Probability
distribution function of dislocation density fluctuations corresponding
to PFC single crystals sheared at different shear rates. The continuous
lines represent the stationary probability distributions predicted by
the stochastic model of dislocation number fluctuations from
Eq.~(\ref{eq:Pd}). Numerical values of the model fitting parameters are
tabulated below.}
\begin{tabular}{|c|c|c|c|c|}
    \hline
   $\dot\epsilon_0$ & $2.78\times 10^{-4}$ & $3.02\times 10^{-4}$ &$3.41\times 10^{4}$ & $3.90\times 10^{-4}$ \\\hline
   $k$ & $0.31$ & $0.62$ & $2.67$ & $3.20$ \\
   $D$ & $0.09$ & $0.09$ & $0.18$ & $0.12$  \\
   $m$ & $26.01$ & $28.08$ & $29.08$ & $31.15$\\\hline
\end{tabular}
\end{figure}

For each constant shear rate $\dot\epsilon_0$, we also measure the
probability distribution function (PDF) of the distribution number as
shown in Figure~(\ref{fig:5}). At low $\dot\epsilon_0$'s, the PDF
$P(N_d)$ is bimodal with one peak near zero corresponding to the
frequency of dislocation extinctions and the other peak centered around
a mean value imposed by the external shear rate. At higher values of
$\dot\epsilon_0$, the probability that the dislocation number drops to
zero becomes vanishingly small, and approaches a uni-modal form with
the peak shifting to the right as the driving force is increased.

{\it Stochastic model:-} The $1/f^{2}$ signal signature leads us to
adopt a stochastic approach for the description of the dynamics of mobile dislocations
initially proposed by H{\"a}hner~\cite{hahner1996foundations,hahner1996stochastic} to
describe dislocation cell formation and fractal patterning during
various plastic deformation regimes.  The basic idea is that the
intrinsic stress and strain-rate fluctuations, arising from long-range
interactions between dislocations, may lead to noise-induced
non-equilibrium phase transitions corresponding to different plastic
deformation regimes and dislocation patterning. The microscopic
details of dislocation interactions are neglected and effectively
replaced by a stochastic contribution to the mean field. Thus, one can
write a general evolution equation of the dislocation density $\rho_d =
N_d/\mathcal{V}$, where $\mathcal{V}$ is the crystal volume in which
dislocations are embedded, as
\begin{equation}\label{eq:rho_d_1}
\dot\rho_d = R(\rho_d,\dot\epsilon),
\end{equation}
where dislocation reactions depend on the strain-rate $\dot\epsilon$
and the defect density $\rho_d$. We decompose the strain-rate into a
deterministic part related to the external driving and a stochastic
part related to dislocation interactions, $\dot\epsilon = \langle
\dot\epsilon\rangle+\delta\dot\epsilon$. The simplest assumption is
that all reactions rate depend linearly on the stochastic strain rate,
so that Eq.~(\ref{eq:rho_d_1}) reduces to a Langevin equation of the
form
\begin{equation}\label{eq:Lagevin}
\dot\rho_d = F(\rho_d, \langle\dot\epsilon\rangle)+G(\rho_d)\delta\dot\epsilon,
\end{equation}
where $F(\rho_d,\langle\dot\epsilon\rangle)$ describes the
deterministic reaction rates depending on the current dislocation
density and the mean strain rate, while $G(\rho_d)$ models the
stochastic reaction rate due to mutual dislocation interactions
depending on the density $\rho_d$. The strain-rate fluctuations
$\delta\dot\epsilon(t)$ are approximated by a Gaussian white noise with
zero mean and covariance $\langle
\delta\dot\epsilon(t)\delta\dot\epsilon(t')\rangle = 2D\delta(t-t')$,
while the noise amplitude $\sqrt{D}$ measures the effective contribution of
dislocation interactions. The white noise-limit is taken under the
approximation that the  strain-rate fluctuations are typically
short-ranged compared to the timescale of dislocation evolution and
patterning~\cite{hahner1996foundations}. The probability distribution
function of dislocation density $P(\rho_d)$ follows from
Eq.~(\ref{eq:Lagevin}) as the steady state solution of the
Fokker-Planck equation in the Stratonovich formulation with natural
boundary conditions and given as
\begin{equation}\label{eq:pdf_1}
P(\rho_d) = \frac{\mathcal{N}}{G(\rho_d)}\exp\left(\int_0^{\rho_d}dx\frac{F(\rho)}{DG^2(\rho)}\right),
\end{equation}
where $\mathcal{N}^{-1} = \int_0^\infty d\rho_d
G^{-1}(\rho_d)\exp\left(\int_0^{\rho_d}d\rho
f(\rho)D^{-1}G^{-2}(\rho)\right)$ is the normalization constant. We
assume that the deterministic part of the dislocation reactions is
driven by a potential field, i.e. $F(\rho_d) = -U'(\rho_d)$, that is
approximated to the lowest order by a double-well potential $U(\rho_d)
=
\frac{1}{4}\left(\frac{\rho_d}{m}-1\right)^4-\frac{1}{2}\left(\frac{\rho_d}{m}-1\right)^2-\kappa\left(\frac{\rho_d}{m}-1\right)$,
with the two minima corresponding to zero density and a mean density
related to the mean strain rate. Thus, the scaling parameter $m$
locates the mean density that increases monotonically with the shear
rate $\dot\epsilon_0$ (also seen from the table in
Figure~(\ref{fig:5})). $\kappa$ is a parameter that also increases with
$\dot\epsilon_0$ and favors a finite mean density. As
$\kappa\rightarrow 0$, dislocation extinction events and a finite
population of interacting dislocations both occur with non-zero
probabilities. The noise intensity in Eq.~(\ref{eq:Lagevin}) depends on
the dislocation density, such that it is able to simulate the internal interactions
between dislocations. To this end, we assume a linear
relationship given by $G(\rho_d) = 1+\rho_d/m$. With these specific
expressions of the reaction rates, we find that the PDF of $\rho_d$
given generically by Eq.~(\ref{eq:pdf_1}) becomes equal to
\begin{equation}\label{eq:Pd}
P(\rho_d) = \mathcal{N} \left(1+\frac{\rho_d}{m}\right)^{-1-11/D}\exp\left(-\frac{\mathcal{L}(\rho_d)}{2D}\right),
\end{equation}
where $\mathcal{L}=
\frac{\rho_d}{m^2(m+\rho_d)}(\rho_d^2-9m\rho_d-2m^2\kappa-22m^2)$. The
stationary distribution of this simple stochastic model of dislocation
reactions captures very well the empirical PDF obtained from the
phase-field crystal simulations as seen in Figure~(\ref{fig:5}). We
have also verified numerically that our stochastic model is able to
reproduce the $1/f^2$ spectral density of number fluctuations. An
analytical calculation of the high-frequency limit of the
power-spectrum using the results of Refs.~\cite{caroli1982high,
brey1984spectral, sigeti1987high} is given in the Supplemental
Material. There we show that the correlation function $C(t) =
\langle\rho_d(0)\rho_d(t)\rangle$ can be calculated in general as a
power series of the form
$C(t)=\sum_{n=0}^\infty\frac{(-t)^n}{n!}\langle
x(\mathcal{O}^\dagger)^n x\rangle$, where $\mathcal{O}^\dagger =
-\left[F+DG'G)\right]\frac{\partial}{\partial
\rho_d}-DG^2\frac{\partial^2}{\partial \rho_d^2}$ is the adjoint
Kolmogorov operator corresponding to Eq.~(\ref{eq:Lagevin}). Hence, the
large $f$ limit of the power spectrum $S(f)= 2 \Re\{\int_0^\infty dt
e^{-2\pi i f t}C(t)\}$ is dominated by the first non-zero term in the
expansion and given as $S(f) \approx \langle \rho_d \mathcal{O}^\dagger
\rho_d\rangle f^{-2}$, where $\langle \rho_d \mathcal{O}^\dagger
\rho_d\rangle = -\langle\rho_d[F+DG'G]\rangle$ is determined by the
first four moments of the steady state $P(\rho_d)$.  Thus, the high frequency power spectrum is proportional to $1/f^2$, with the proportionality constant depending on the mean density and higher moments, a result that is expected to be
robust in higher dimensions and testable in experiment.

In conclusion, by using the phase field crystal model, we showed that
number fluctuations of dislocations follow non-Gaussian statistics with
a $1/f^2$ power spectrum similar to that of strain rate fluctuations.
This behavior arises from correlated dislocation reactions, and can be
accurately captured by a stochastic model which makes experimentally
testable predictions for the probability distribution of defect numbers
as a function of shear rate.

\bibliographystyle{apsrev4-1}
\bibliography{refs}

\newpage
\centerline{\bf Supplementary Information}

The dislocation density fluctuations are modelled by a non-linear Langevin equation with multiplicative noise given generically as
\begin{equation}\label{eq:lagevin}
\dot x = f(x)+g(x)\xi(t),
\end{equation}
where the noise term is related to the strain-rate fluctuations due to dislocation interactions, and $\xi(t)$ is approximated by a Gaussian white noise with zero mean
\begin{equation}\label{eq:xi}
\langle \xi(t)\xi(t')\rangle = 2D\delta(t-t'),
\end{equation}
with the $D$ a parameter of noise amplitude that measures the effective contribution of dislocation interactions. The deterministic part $f(x)$ describes the dislocation reaction rate and it is driven by a potential field, i.e. $f(x) = -dU(x)/dx$, that is approximated to the lowest order by a double-well potential

\begin{equation}
U(x) = \frac{1}{4}\left(\frac{x}{m}-1\right)^4-\frac{1}{2}\left(\frac{x}{m}-1\right)^2-\kappa\left(\frac{x}{m}-1\right),
\end{equation}

with the two minima corresponding to zero density and a mean density related to the mean strain rate. Thus, the scaling parameter $m$ locates the mean density that increases monotonically with the shear rate $\dot\epsilon_0$. $\kappa$ is a parameter that also increases with $\dot\epsilon_0$ and favours a finite mean density. As $\kappa\rightarrow 0$, dislocation extinction events and a finite population of interacting dislocations both occur with non-zero probabilities. The noise intensity in Eq.~(\ref{eq:lagevin}) depends on the dislocation density, such that it is able to simulate the internal interactions between dislocations. To this end, we assume a linear relationship given by $g(x) = 1+x/m$.

For a generic Langevin equation with additive noise, there is a general argument to determine the high-frequency limit of the noise power-spectrum~\cite{caroli1982high, brey1984spectral, sigeti1987high}. One can show that the power spectrum density $S(\omega)$ of one-dimensional stochastic fluctuations decays asymptotically as $1/\omega^2$ for additive Gaussian noise regardless of expression of the deterministic drive~\cite{caroli1982high, brey1984spectral}, whereas for a second-order, additive Langevin equation, $S(\omega)$ is dominated by $1/\omega^4$ as $\omega\rightarrow\infty$. Here, we present a general derivation of $S(\omega)$ in the high-$\omega$ limit for a fist-order, multiplicative Langevin equation given by Eq.~(\ref{eq:lagevin}).

The multiplicative noise in Eq.~(\ref{eq:lagevin}) is studied using the Stratonovich calculus when the strain-rate fluctuations are considered in the $\delta$-correlated limit. Henceforth, the corresponding Fokker-Planck equation is given by
\begin{equation}\label{eq:FP}
\frac{\partial P(x,t)}{\partial t} = -\mathcal{O}P(x,t),
\end{equation}
where $P(x,t)$ is the probability distribution function and the propagation operator $\mathcal{O}$ in the Stratonovich definition  takes the form
\begin{equation}
\mathcal{O} = \frac{\partial}{\partial x}\left[f(x)+Dg'(x)g(x)\right]-D\frac{\partial^2}{\partial x^2}g^2(x).
\end{equation}
Similar results are also obtained in the Ito calculus for which the $g$-dependent term in the drift term is removed.

The formal solution of Eq.~(\ref{eq:FP}) with the initial condition $P(x_0,0) = \delta(x-x_0)$ can be expressed as
\begin{equation}\label{eq:Pxt}
P(x,t|x_0,0) = e^{-\mathcal{O}t}\delta(x-x_0).
\end{equation}
Also the stationary probability distribution $P_{st}(x)$ of Eq.~(\ref{eq:FP}) is
\begin{equation}\label{eq:pdf_1}
P_{st}(x) = \frac{\mathcal{N}}{g(x)}\exp\left(\int_0^{x}dy\frac{f(y)}{Dg^2(y)}\right),
\end{equation}
where $\mathcal{N}^{-1} = \int_0^\infty dx g^{-1}(x)\exp\left(\int_0^{x}dyf(y)D^{-1}g^{-2}(y)\right)$ is the normalization constant.  With the specific expressions of the reaction rates that we considered for our model, we find that the PDF of $x$ given generically by Eq.~(\ref{eq:pdf_1}) becomes equal to
\begin{equation}\label{eq:Pd}
P_{st}(x) = \mathcal{N} \left(1+\frac{x}{m}\right)^{-1-11/D}\exp\left(-\frac{\mathcal{L}(x)}{2D}\right),
\end{equation}
where $\mathcal{L}= \frac{x}{m^2(m+x)}(x^2-9mx-2m^2\kappa-22m^2)$.

The power spectrum density $S(\omega)$ is defined as the Fourier transform of the time correlation function and given as
\begin{equation}\label{eq:Sf}
S(\omega) = \frac{1}{\pi}Re\int_0^\infty d\omega e^{-i\omega t}C(t),
\end{equation}
where the time correlation function is
\begin{eqnarray}
C(t) &=& \langle x(t)x(0)\rangle \nonumber\\
&=& \int dx\int dx_0 xP(x,t|x_0,0)x_0 P_{st}(x_0).
\end{eqnarray}
Using the formal solution of the transition probability from Eq.~(\ref{eq:Pxt}), we can integrate over $x_0$ and arrive at
\begin{eqnarray}
C(t) &=& \int dx x e^{-\mathcal{O}t}x P_{st}(x)\nonumber\\
&=& \int dx P_{st}(x) \left(x e^{-\mathcal{O}^\dagger t}x\right)  = \langle xe^{-\mathcal{O}^\dagger t}x\rangle,\label{eq:Ct}
\end{eqnarray}
where $\langle\cdot\rangle$ represents an average with respect to the stationary distribution $P_{st}$. The adjoint operator $\mathcal{O}^\dagger$ is defined from the relation $\int dx \psi(x)\mathcal{O}\phi(x) = \int dx \phi(x)\mathcal{O}^\dagger\psi(x)$ and after integration by parts is given as
\begin{equation}
\mathcal{O}^\dagger = -\left[f(x)+Dg'(x)g(x)\right]\frac{\partial}{\partial x}-Dg^2(x)\frac{\partial^2}{\partial x^2}.
\end{equation}
Expanding the exponential in Eq.~(\ref{eq:Ct}), we have that
\begin{eqnarray}
C(t) &=& \sum_{n=0}^\infty\frac{(-t)^n}{n!}\langle x(\mathcal{O}^\dagger)^n x\rangle,
\end{eqnarray}
and the power spectrum from Eq.~(\ref{eq:Sf}) becomes equal to
\begin{eqnarray}\label{eq:Sf_series}
S(\omega) &=& \frac{1}{\pi}Re \sum_{n=0}^\infty \frac{(-1)^n}{n!}\langle x(\mathcal{O}^\dagger)^n x\rangle\int_0^\infty d\omega e^{-i\omega t}t^n\nonumber\\
&=&-\frac{1}{\pi}Re \sum_{n=0}^\infty \frac{\langle x(\mathcal{O}^\dagger)^n x\rangle}{(-i\omega)^{n+1}}.
\end{eqnarray}
In the high-$\omega$ limit, the power spectrum density is dominated by the first non-zero term in the series given that the expansion coefficients are well-defined and decrease in amplitude with increasing order. For our definitions of $f(x)$ and $g(x)$, we have that $|\langle x(\mathcal{O}^\dagger)^n x\rangle|>|\langle x(\mathcal{O}^\dagger)^{n+1} x\rangle|$ for all $n\ge 1$. The zero order term vanishes since it is purely imaginary, hence the first non-zero term corresponds to $n=1$,
\begin{eqnarray}
S(\omega) \approx \frac{1}{\pi}\frac{\langle x \mathcal{O}^\dagger x\rangle}{\omega^2},
\end{eqnarray}
where $x \mathcal{O}^\dagger x = -x[f(x)+Dg'(x)g(x)]$ or $x \mathcal{O}^\dagger x = -(Dm^3+Dm^2x+km^3-2m^2x+3mx^2-x^3)/m^4$. Thus, the power-spectrum density of the fluctuations described by Eq.~(\ref{eq:lagevin}) decays asymptotically as $1/\omega^2$ when $\langle x \mathcal{O}^\dagger x\rangle$ is not identically zero.  We have also checked this numerically for the specific expressions of $f(x)$ and $g(x)$.

\bibliographystyle{apsrev4-1}
\bibliography{refs}

\begin{thebibliography}{25}%
\makeatletter
\providecommand \@ifxundefined [1]{%
 \@ifx{#1\undefined}
}%
\providecommand \@ifnum [1]{%
 \ifnum #1\expandafter \@firstoftwo
 \else \expandafter \@secondoftwo
 \fi
}%
\providecommand \@ifx [1]{%
 \ifx #1\expandafter \@firstoftwo
 \else \expandafter \@secondoftwo
 \fi
}%
\providecommand \natexlab [1]{#1}%
\providecommand \enquote  [1]{``#1''}%
\providecommand \bibnamefont  [1]{#1}%
\providecommand \bibfnamefont [1]{#1}%
\providecommand \citenamefont [1]{#1}%
\providecommand \href@noop [0]{\@secondoftwo}%
\providecommand \href [0]{\begingroup \@sanitize@url \@href}%
\providecommand \@href[1]{\@@startlink{#1}\@@href}%
\providecommand \@@href[1]{\endgroup#1\@@endlink}%
\providecommand \@sanitize@url [0]{\catcode `\\12\catcode `\$12\catcode
  `\&12\catcode `\#12\catcode `\^12\catcode `\_12\catcode `\%12\relax}%
\providecommand \@@startlink[1]{}%
\providecommand \@@endlink[0]{}%
\providecommand \url  [0]{\begingroup\@sanitize@url \@url }%
\providecommand \@url [1]{\endgroup\@href {#1}{\urlprefix }}%
\providecommand \urlprefix  [0]{URL }%
\providecommand \Eprint [0]{\href }%
\providecommand \doibase [0]{http://dx.doi.org/}%
\providecommand \selectlanguage [0]{\@gobble}%
\providecommand \bibinfo  [0]{\@secondoftwo}%
\providecommand \bibfield  [0]{\@secondoftwo}%
\providecommand \translation [1]{[#1]}%
\providecommand \BibitemOpen [0]{}%
\providecommand \bibitemStop [0]{}%
\providecommand \bibitemNoStop [0]{.\EOS\space}%
\providecommand \EOS [0]{\spacefactor3000\relax}%
\providecommand \BibitemShut  [1]{\csname bibitem#1\endcsname}%
\let\auto@bib@innerbib\@empty
\bibitem [{\citenamefont {Richeton}\ \emph {et~al.}(2005)\citenamefont
  {Richeton}, \citenamefont {Weiss},\ and\ \citenamefont
  {Louchet}}]{richeton2005dislocation}%
  \BibitemOpen
  \bibfield  {author} {\bibinfo {author} {\bibfnamefont {T.}~\bibnamefont
  {Richeton}}, \bibinfo {author} {\bibfnamefont {J.}~\bibnamefont {Weiss}}, \
  and\ \bibinfo {author} {\bibfnamefont {F.}~\bibnamefont {Louchet}},\
  }\href@noop {} {\bibfield  {journal} {\bibinfo  {journal} {Acta materialia}\
  }\textbf {\bibinfo {volume} {53}},\ \bibinfo {pages} {4463} (\bibinfo {year}
  {2005})}\BibitemShut {NoStop}%
\bibitem [{\citenamefont {Richeton}\ \emph {et~al.}(2006)\citenamefont
  {Richeton}, \citenamefont {Dobron}, \citenamefont {Chmelik}, \citenamefont
  {Weiss},\ and\ \citenamefont {Louchet}}]{richeton2006critical}%
  \BibitemOpen
  \bibfield  {author} {\bibinfo {author} {\bibfnamefont {T.}~\bibnamefont
  {Richeton}}, \bibinfo {author} {\bibfnamefont {P.}~\bibnamefont {Dobron}},
  \bibinfo {author} {\bibfnamefont {F.}~\bibnamefont {Chmelik}}, \bibinfo
  {author} {\bibfnamefont {J.}~\bibnamefont {Weiss}}, \ and\ \bibinfo {author}
  {\bibfnamefont {F.}~\bibnamefont {Louchet}},\ }\href@noop {} {\bibfield
  {journal} {\bibinfo  {journal} {Materials Science and Engineering: A}\
  }\textbf {\bibinfo {volume} {424}},\ \bibinfo {pages} {190} (\bibinfo {year}
  {2006})}\BibitemShut {NoStop}%
\bibitem [{\citenamefont {Greer}\ \emph {et~al.}(2009)\citenamefont {Greer},
  \citenamefont {Kim},\ and\ \citenamefont {Burek}}]{greer2009situ}%
  \BibitemOpen
  \bibfield  {author} {\bibinfo {author} {\bibfnamefont {J.~R.}\ \bibnamefont
  {Greer}}, \bibinfo {author} {\bibfnamefont {J.-Y.}\ \bibnamefont {Kim}}, \
  and\ \bibinfo {author} {\bibfnamefont {M.~J.}\ \bibnamefont {Burek}},\
  }\href@noop {} {\bibfield  {journal} {\bibinfo  {journal} {JOM}\ }\textbf
  {\bibinfo {volume} {61}},\ \bibinfo {pages} {19} (\bibinfo {year}
  {2009})}\BibitemShut {NoStop}%
\bibitem [{\citenamefont {Argon}(2013)}]{argon2013strain}%
  \BibitemOpen
  \bibfield  {author} {\bibinfo {author} {\bibfnamefont {A.}~\bibnamefont
  {Argon}},\ }\href@noop {} {\bibfield  {journal} {\bibinfo  {journal}
  {Philosophical Magazine}\ }\textbf {\bibinfo {volume} {93}},\ \bibinfo
  {pages} {3795} (\bibinfo {year} {2013})}\BibitemShut {NoStop}%
\bibitem [{\citenamefont {Miguel}\ \emph {et~al.}(2001)\citenamefont {Miguel},
  \citenamefont {Vespignani}, \citenamefont {Zapperi}, \citenamefont {Weiss},\
  and\ \citenamefont {Grasso}}]{miguel2001intermittent}%
  \BibitemOpen
  \bibfield  {author} {\bibinfo {author} {\bibfnamefont {M.-C.}\ \bibnamefont
  {Miguel}}, \bibinfo {author} {\bibfnamefont {A.}~\bibnamefont {Vespignani}},
  \bibinfo {author} {\bibfnamefont {S.}~\bibnamefont {Zapperi}}, \bibinfo
  {author} {\bibfnamefont {J.}~\bibnamefont {Weiss}}, \ and\ \bibinfo {author}
  {\bibfnamefont {J.-R.}\ \bibnamefont {Grasso}},\ }\href@noop {} {\bibfield
  {journal} {\bibinfo  {journal} {Nature}\ }\textbf {\bibinfo {volume} {410}},\
  \bibinfo {pages} {667} (\bibinfo {year} {2001})}\BibitemShut {NoStop}%
\bibitem [{\citenamefont {Isp{\'a}novity}\ \emph {et~al.}(2010)\citenamefont
  {Isp{\'a}novity}, \citenamefont {Groma}, \citenamefont {Gy{\"o}rgyi},
  \citenamefont {Csikor},\ and\ \citenamefont
  {Weygand}}]{ispanovity2010submicron}%
  \BibitemOpen
  \bibfield  {author} {\bibinfo {author} {\bibfnamefont {P.~D.}\ \bibnamefont
  {Isp{\'a}novity}}, \bibinfo {author} {\bibfnamefont {I.}~\bibnamefont
  {Groma}}, \bibinfo {author} {\bibfnamefont {G.}~\bibnamefont {Gy{\"o}rgyi}},
  \bibinfo {author} {\bibfnamefont {F.~F.}\ \bibnamefont {Csikor}}, \ and\
  \bibinfo {author} {\bibfnamefont {D.}~\bibnamefont {Weygand}},\ }\href@noop
  {} {\bibfield  {journal} {\bibinfo  {journal} {Physical Review lLtters}\
  }\textbf {\bibinfo {volume} {105}},\ \bibinfo {pages} {085503} (\bibinfo
  {year} {2010})}\BibitemShut {NoStop}%
\bibitem [{\citenamefont {Tsekenis}\ \emph {et~al.}(2013)\citenamefont
  {Tsekenis}, \citenamefont {Uhl}, \citenamefont {Goldenfeld},\ and\
  \citenamefont {Dahmen}}]{tsekenis2013determination}%
  \BibitemOpen
  \bibfield  {author} {\bibinfo {author} {\bibfnamefont {G.}~\bibnamefont
  {Tsekenis}}, \bibinfo {author} {\bibfnamefont {J.}~\bibnamefont {Uhl}},
  \bibinfo {author} {\bibfnamefont {N.}~\bibnamefont {Goldenfeld}}, \ and\
  \bibinfo {author} {\bibfnamefont {K.}~\bibnamefont {Dahmen}},\ }\href@noop {}
  {\bibfield  {journal} {\bibinfo  {journal} {EPL (Europhysics Letters)}\
  }\textbf {\bibinfo {volume} {101}},\ \bibinfo {pages} {36003} (\bibinfo
  {year} {2013})}\BibitemShut {NoStop}%
\bibitem [{\citenamefont {Friedman}\ \emph {et~al.}(2012)\citenamefont
  {Friedman}, \citenamefont {Jennings}, \citenamefont {Tsekenis}, \citenamefont
  {Kim}, \citenamefont {Tao}, \citenamefont {Uhl}, \citenamefont {Greer},\ and\
  \citenamefont {Dahmen}}]{friedman2012statistics}%
  \BibitemOpen
  \bibfield  {author} {\bibinfo {author} {\bibfnamefont {N.}~\bibnamefont
  {Friedman}}, \bibinfo {author} {\bibfnamefont {A.~T.}\ \bibnamefont
  {Jennings}}, \bibinfo {author} {\bibfnamefont {G.}~\bibnamefont {Tsekenis}},
  \bibinfo {author} {\bibfnamefont {J.-Y.}\ \bibnamefont {Kim}}, \bibinfo
  {author} {\bibfnamefont {M.}~\bibnamefont {Tao}}, \bibinfo {author}
  {\bibfnamefont {J.~T.}\ \bibnamefont {Uhl}}, \bibinfo {author} {\bibfnamefont
  {J.~R.}\ \bibnamefont {Greer}}, \ and\ \bibinfo {author} {\bibfnamefont
  {K.~A.}\ \bibnamefont {Dahmen}},\ }\href@noop {} {\bibfield  {journal}
  {\bibinfo  {journal} {Physical Review Letters}\ }\textbf {\bibinfo {volume}
  {109}},\ \bibinfo {pages} {095507} (\bibinfo {year} {2012})}\BibitemShut
  {NoStop}%
\bibitem [{\citenamefont {LeBlanc}\ \emph {et~al.}(2012)\citenamefont
  {LeBlanc}, \citenamefont {Angheluta}, \citenamefont {Dahmen},\ and\
  \citenamefont {Goldenfeld}}]{leblanc2012distribution}%
  \BibitemOpen
  \bibfield  {author} {\bibinfo {author} {\bibfnamefont {M.}~\bibnamefont
  {LeBlanc}}, \bibinfo {author} {\bibfnamefont {L.}~\bibnamefont {Angheluta}},
  \bibinfo {author} {\bibfnamefont {K.}~\bibnamefont {Dahmen}}, \ and\ \bibinfo
  {author} {\bibfnamefont {N.}~\bibnamefont {Goldenfeld}},\ }\href@noop {}
  {\bibfield  {journal} {\bibinfo  {journal} {Physical Review Letters}\
  }\textbf {\bibinfo {volume} {109}},\ \bibinfo {pages} {105702} (\bibinfo
  {year} {2012})}\BibitemShut {NoStop}%
\bibitem [{\citenamefont {Isp{\'a}novity}\ \emph {et~al.}(2014)\citenamefont
  {Isp{\'a}novity}, \citenamefont {Laurson}, \citenamefont {Zaiser},
  \citenamefont {Groma}, \citenamefont {Zapperi},\ and\ \citenamefont
  {Alava}}]{ispanovity2014avalanches}%
  \BibitemOpen
  \bibfield  {author} {\bibinfo {author} {\bibfnamefont {P.~D.}\ \bibnamefont
  {Isp{\'a}novity}}, \bibinfo {author} {\bibfnamefont {L.}~\bibnamefont
  {Laurson}}, \bibinfo {author} {\bibfnamefont {M.}~\bibnamefont {Zaiser}},
  \bibinfo {author} {\bibfnamefont {I.}~\bibnamefont {Groma}}, \bibinfo
  {author} {\bibfnamefont {S.}~\bibnamefont {Zapperi}}, \ and\ \bibinfo
  {author} {\bibfnamefont {M.~J.}\ \bibnamefont {Alava}},\ }\href@noop {}
  {\bibfield  {journal} {\bibinfo  {journal} {Physical Review Letters}\
  }\textbf {\bibinfo {volume} {112}},\ \bibinfo {pages} {235501} (\bibinfo
  {year} {2014})}\BibitemShut {NoStop}%
\bibitem [{\citenamefont {Isp{\'a}novity}\ \emph {et~al.}(2013)\citenamefont
  {Isp{\'a}novity}, \citenamefont {Hegyi}, \citenamefont {Groma}, \citenamefont
  {Gy{\"o}rgyi}, \citenamefont {Ratter},\ and\ \citenamefont
  {Weygand}}]{ispanovity2013average}%
  \BibitemOpen
  \bibfield  {author} {\bibinfo {author} {\bibfnamefont {P.~D.}\ \bibnamefont
  {Isp{\'a}novity}}, \bibinfo {author} {\bibfnamefont {{\'A}.}~\bibnamefont
  {Hegyi}}, \bibinfo {author} {\bibfnamefont {I.}~\bibnamefont {Groma}},
  \bibinfo {author} {\bibfnamefont {G.}~\bibnamefont {Gy{\"o}rgyi}}, \bibinfo
  {author} {\bibfnamefont {K.}~\bibnamefont {Ratter}}, \ and\ \bibinfo {author}
  {\bibfnamefont {D.}~\bibnamefont {Weygand}},\ }\href@noop {} {\bibfield
  {journal} {\bibinfo  {journal} {Acta Materialia}\ }\textbf {\bibinfo {volume}
  {61}},\ \bibinfo {pages} {6234} (\bibinfo {year} {2013})}\BibitemShut
  {NoStop}%
\bibitem [{\citenamefont {Elder}\ and\ \citenamefont {Grant}(2004)}]{Elder04}%
  \BibitemOpen
  \bibfield  {author} {\bibinfo {author} {\bibfnamefont {K.}~\bibnamefont
  {Elder}}\ and\ \bibinfo {author} {\bibfnamefont {M.}~\bibnamefont {Grant}},\
  }\href@noop {} {\bibfield  {journal} {\bibinfo  {journal} {Physical Review
  E}\ }\textbf {\bibinfo {volume} {70}},\ \bibinfo {pages} {051605} (\bibinfo
  {year} {2004})}\BibitemShut {NoStop}%
\bibitem [{\citenamefont {Emmerich}\ \emph {et~al.}(2012)\citenamefont
  {Emmerich}, \citenamefont {L{\"o}wen}, \citenamefont {Wittkowski},
  \citenamefont {Gruhn}, \citenamefont {T{\'o}th}, \citenamefont {Tegze},\ and\
  \citenamefont {Gr{\'a}n{\'a}sy}}]{emmerich2012phase}%
  \BibitemOpen
  \bibfield  {author} {\bibinfo {author} {\bibfnamefont {H.}~\bibnamefont
  {Emmerich}}, \bibinfo {author} {\bibfnamefont {H.}~\bibnamefont {L{\"o}wen}},
  \bibinfo {author} {\bibfnamefont {R.}~\bibnamefont {Wittkowski}}, \bibinfo
  {author} {\bibfnamefont {T.}~\bibnamefont {Gruhn}}, \bibinfo {author}
  {\bibfnamefont {G.~I.}\ \bibnamefont {T{\'o}th}}, \bibinfo {author}
  {\bibfnamefont {G.}~\bibnamefont {Tegze}}, \ and\ \bibinfo {author}
  {\bibfnamefont {L.}~\bibnamefont {Gr{\'a}n{\'a}sy}},\ }\href@noop {}
  {\bibfield  {journal} {\bibinfo  {journal} {Advances in Physics}\ }\textbf
  {\bibinfo {volume} {61}},\ \bibinfo {pages} {665} (\bibinfo {year}
  {2012})}\BibitemShut {NoStop}%
\bibitem [{\citenamefont {Berry}\ \emph {et~al.}(2014)\citenamefont {Berry},
  \citenamefont {Provatas}, \citenamefont {Rottler},\ and\ \citenamefont
  {Sinclair}}]{berry2014phase}%
  \BibitemOpen
  \bibfield  {author} {\bibinfo {author} {\bibfnamefont {J.}~\bibnamefont
  {Berry}}, \bibinfo {author} {\bibfnamefont {N.}~\bibnamefont {Provatas}},
  \bibinfo {author} {\bibfnamefont {J.}~\bibnamefont {Rottler}}, \ and\
  \bibinfo {author} {\bibfnamefont {C.~W.}\ \bibnamefont {Sinclair}},\
  }\href@noop {} {\bibfield  {journal} {\bibinfo  {journal} {Physical Review
  B}\ }\textbf {\bibinfo {volume} {89}},\ \bibinfo {pages} {214117} (\bibinfo
  {year} {2014})}\BibitemShut {NoStop}%
\bibitem [{\citenamefont
  {H{\"a}hner}(1996{\natexlab{a}})}]{hahner1996foundations}%
  \BibitemOpen
  \bibfield  {author} {\bibinfo {author} {\bibfnamefont {P.}~\bibnamefont
  {H{\"a}hner}},\ }\href@noop {} {\bibfield  {journal} {\bibinfo  {journal}
  {Applied Physics A}\ }\textbf {\bibinfo {volume} {62}},\ \bibinfo {pages}
  {473} (\bibinfo {year} {1996}{\natexlab{a}})}\BibitemShut {NoStop}%
\bibitem [{\citenamefont
  {H{\"a}hner}(1996{\natexlab{b}})}]{hahner1996stochastic}%
  \BibitemOpen
  \bibfield  {author} {\bibinfo {author} {\bibfnamefont {P.}~\bibnamefont
  {H{\"a}hner}},\ }\href@noop {} {\bibfield  {journal} {\bibinfo  {journal}
  {Applied Physics A}\ }\textbf {\bibinfo {volume} {63}},\ \bibinfo {pages}
  {45} (\bibinfo {year} {1996}{\natexlab{b}})}\BibitemShut {NoStop}%
\bibitem [{\citenamefont {Ananthakrishna}(2007)}]{ananthakrishna2007current}%
  \BibitemOpen
  \bibfield  {author} {\bibinfo {author} {\bibfnamefont {G.}~\bibnamefont
  {Ananthakrishna}},\ }\href@noop {} {\bibfield  {journal} {\bibinfo  {journal}
  {Physics Reports}\ }\textbf {\bibinfo {volume} {440}},\ \bibinfo {pages}
  {113} (\bibinfo {year} {2007})}\BibitemShut {NoStop}%
\bibitem [{\citenamefont {Stefanovic}\ \emph {et~al.}(2006)\citenamefont
  {Stefanovic}, \citenamefont {Haataja},\ and\ \citenamefont
  {Provatas}}]{stefanovic2006phase}%
  \BibitemOpen
  \bibfield  {author} {\bibinfo {author} {\bibfnamefont {P.}~\bibnamefont
  {Stefanovic}}, \bibinfo {author} {\bibfnamefont {M.}~\bibnamefont {Haataja}},
  \ and\ \bibinfo {author} {\bibfnamefont {N.}~\bibnamefont {Provatas}},\
  }\href@noop {} {\bibfield  {journal} {\bibinfo  {journal} {Physical Review
  Letters}\ }\textbf {\bibinfo {volume} {96}},\ \bibinfo {pages} {225504}
  (\bibinfo {year} {2006})}\BibitemShut {NoStop}%
\bibitem [{\citenamefont {Stefanovic}\ \emph {et~al.}(2009)\citenamefont
  {Stefanovic}, \citenamefont {Haataja},\ and\ \citenamefont
  {Provatas}}]{Stefanovic09}%
  \BibitemOpen
  \bibfield  {author} {\bibinfo {author} {\bibfnamefont {P.}~\bibnamefont
  {Stefanovic}}, \bibinfo {author} {\bibfnamefont {M.}~\bibnamefont {Haataja}},
  \ and\ \bibinfo {author} {\bibfnamefont {N.}~\bibnamefont {Provatas}},\
  }\href@noop {} {\bibfield  {journal} {\bibinfo  {journal} {Physical Review
  E}\ }\textbf {\bibinfo {volume} {80}},\ \bibinfo {pages} {046107} (\bibinfo
  {year} {2009})}\BibitemShut {NoStop}%
\bibitem [{\citenamefont {Chan}\ \emph {et~al.}(2010)\citenamefont {Chan},
  \citenamefont {Tsekenis}, \citenamefont {Dantzig}, \citenamefont {Dahmen},\
  and\ \citenamefont {Goldenfeld}}]{Chan10}%
  \BibitemOpen
  \bibfield  {author} {\bibinfo {author} {\bibfnamefont {P.~Y.}\ \bibnamefont
  {Chan}}, \bibinfo {author} {\bibfnamefont {G.}~\bibnamefont {Tsekenis}},
  \bibinfo {author} {\bibfnamefont {J.}~\bibnamefont {Dantzig}}, \bibinfo
  {author} {\bibfnamefont {K.~A.}\ \bibnamefont {Dahmen}}, \ and\ \bibinfo
  {author} {\bibfnamefont {N.}~\bibnamefont {Goldenfeld}},\ }\href@noop {}
  {\bibfield  {journal} {\bibinfo  {journal} {Physical Review Letters}\
  }\textbf {\bibinfo {volume} {105}},\ \bibinfo {pages} {015502} (\bibinfo
  {year} {2010})}\BibitemShut {NoStop}%
\bibitem [{\citenamefont {Angheluta}\ \emph {et~al.}(2012)\citenamefont
  {Angheluta}, \citenamefont {Jeraldo},\ and\ \citenamefont
  {Goldenfeld}}]{PRE12}%
  \BibitemOpen
  \bibfield  {author} {\bibinfo {author} {\bibfnamefont {L.}~\bibnamefont
  {Angheluta}}, \bibinfo {author} {\bibfnamefont {P.}~\bibnamefont {Jeraldo}},
  \ and\ \bibinfo {author} {\bibfnamefont {N.}~\bibnamefont {Goldenfeld}},\
  }\href@noop {} {\bibfield  {journal} {\bibinfo  {journal} {Physical Review
  E}\ }\textbf {\bibinfo {volume} {85}},\ \bibinfo {pages} {011153} (\bibinfo
  {year} {2012})}\BibitemShut {NoStop}%
\bibitem [{\citenamefont {Tegze}\ \emph {et~al.}(2009)\citenamefont {Tegze},
  \citenamefont {Bansel}, \citenamefont {T{\'o}th}, \citenamefont {Pusztai},
  \citenamefont {Fan},\ and\ \citenamefont
  {Gr{\'a}n{\'a}sy}}]{tegze2009advanced}%
  \BibitemOpen
  \bibfield  {author} {\bibinfo {author} {\bibfnamefont {G.}~\bibnamefont
  {Tegze}}, \bibinfo {author} {\bibfnamefont {G.}~\bibnamefont {Bansel}},
  \bibinfo {author} {\bibfnamefont {G.~I.}\ \bibnamefont {T{\'o}th}}, \bibinfo
  {author} {\bibfnamefont {T.}~\bibnamefont {Pusztai}}, \bibinfo {author}
  {\bibfnamefont {Z.}~\bibnamefont {Fan}}, \ and\ \bibinfo {author}
  {\bibfnamefont {L.}~\bibnamefont {Gr{\'a}n{\'a}sy}},\ }\href@noop {}
  {\bibfield  {journal} {\bibinfo  {journal} {Journal of Computational
  Physics}\ }\textbf {\bibinfo {volume} {228}},\ \bibinfo {pages} {1612}
  (\bibinfo {year} {2009})}\BibitemShut {NoStop}%
\bibitem [{\citenamefont {Caroli}\ \emph {et~al.}(1982)\citenamefont {Caroli},
  \citenamefont {Caroli},\ and\ \citenamefont {Roulet}}]{caroli1982high}%
  \BibitemOpen
  \bibfield  {author} {\bibinfo {author} {\bibfnamefont {B.}~\bibnamefont
  {Caroli}}, \bibinfo {author} {\bibfnamefont {C.}~\bibnamefont {Caroli}}, \
  and\ \bibinfo {author} {\bibfnamefont {B.}~\bibnamefont {Roulet}},\
  }\href@noop {} {\bibfield  {journal} {\bibinfo  {journal} {Physica A:
  Statistical Mechanics and its Applications}\ }\textbf {\bibinfo {volume}
  {112}},\ \bibinfo {pages} {517} (\bibinfo {year} {1982})}\BibitemShut
  {NoStop}%
\bibitem [{\citenamefont {Brey}\ \emph {et~al.}(1984)\citenamefont {Brey},
  \citenamefont {Casado},\ and\ \citenamefont {Morillo}}]{brey1984spectral}%
  \BibitemOpen
  \bibfield  {author} {\bibinfo {author} {\bibfnamefont {J.}~\bibnamefont
  {Brey}}, \bibinfo {author} {\bibfnamefont {J.}~\bibnamefont {Casado}}, \ and\
  \bibinfo {author} {\bibfnamefont {M.}~\bibnamefont {Morillo}},\ }\href@noop
  {} {\bibfield  {journal} {\bibinfo  {journal} {Physical Review A}\ }\textbf
  {\bibinfo {volume} {30}},\ \bibinfo {pages} {1535} (\bibinfo {year}
  {1984})}\BibitemShut {NoStop}%
\bibitem [{\citenamefont {Sigeti}\ and\ \citenamefont
  {Horsthemke}(1987)}]{sigeti1987high}%
  \BibitemOpen
  \bibfield  {author} {\bibinfo {author} {\bibfnamefont {D.}~\bibnamefont
  {Sigeti}}\ and\ \bibinfo {author} {\bibfnamefont {W.}~\bibnamefont
  {Horsthemke}},\ }\href@noop {} {\bibfield  {journal} {\bibinfo  {journal}
  {Physical Review A}\ }\textbf {\bibinfo {volume} {35}},\ \bibinfo {pages}
  {2276} (\bibinfo {year} {1987})}\BibitemShut {NoStop}%
\end{thebibliography}%

\end{document}